\documentclass{article}

\usepackage{arxiv}

\usepackage[utf8]{inputenc} 
\usepackage[T1]{fontenc}    
\usepackage{url}            
\usepackage{booktabs}       
\usepackage{amsfonts}       
\usepackage{nicefrac}       
\usepackage{microtype}      
\usepackage{lipsum}
\usepackage[numbers,sort]{natbib}

\usepackage[english]{babel}
\usepackage{float}
\usepackage{amsmath}
\usepackage{amsfonts}
\usepackage{graphicx}
\usepackage{algorithm}
\usepackage{algpseudocode}
\usepackage[colorinlistoftodos]{todonotes}
\usepackage[colorlinks=true, allcolors=blue]{hyperref}

\title{AITom: Open-source AI platform for cryo-electron Tomography data analysis}

\author{
  Xiangrui Zeng \\
  Computational Biology Department\\
  School of Computer Science \\
  Carnegie Mellon University\\
  Pittsburgh, PA 15213 \\
  \texttt{xiangruz@andrew.cmu.edu} \\
   \And
  Min Xu\thanks{Corresponding author} \\
  Computational Biology Department\\
  School of Computer Science \\
  Carnegie Mellon University\\
  Pittsburgh, PA 15213 \\
  \texttt{mxu1@cs.cmu.edu} \\
}

\begin{document}

\maketitle

\begin{abstract}
Cryo-electron tomography (cryo-ET) is an emerging technology for the 3D visualization of structural organizations and interactions of subcellular components at near-native state and sub-molecular resolution. Tomograms captured by cryo-ET contain heterogeneous structures representing the complex and dynamic subcellular environment. Since the structures are not purified or fluorescently labeled, the spatial organization and interaction between both the known and unknown structures can be studied in their native environment. 
The rapid advances of cryo-electron tomography (cryo-ET) have generated abundant 3D cellular imaging data. However, the systematic localization, identification, segmentation, and structural recovery of the subcellular components require efficient and accurate large-scale image analysis methods. We introduce AITom, an open-source artificial intelligence platform for cryo-ET researchers. AITom provides many public as well as in-house algorithms for performing cryo-ET data analysis through both the traditional template-based or template-free approach and the deep learning approach. AITom also supports remote interactive analysis. Comprehensive tutorials for each analysis module are provided to guide the user through. We welcome researchers and developers to join this collaborative open-source software development project.

\textbf{Availability}: \url{https://github.com/xulabs/aitom}

\end{abstract}


\section{Introduction}
Cryo-electron microscopy is a revolutionary technique that determines the three-dimensional structure of macromolecular complexes at nano-resolution \cite{luvcic2013cryo}. The cryo-electron microscopy can also be used to produce 2D projection images from a cell sample. The 3D image can then be reconstructed from the set of 2D projection images through a tilt angle series. This technique is referred to as cryo-electron tomography (cryo-ET). The reconstructed 3D image is called a tomogram. The tomogram preserves the native structure and spatial organizations of macromolecular complexes and cellular ultrastructure (such as membranes) in the cytoplasm \cite{oikonomou2017cellular}. However, major gaps exist between obtaining the raw tomogram data and decoding the macromolecular complex structural information from it. 

A major problem with the analysis of tomogram data is image distortion. The main image distortion is noise. Due to the thickness of the cell sample and the crowded cellular environment, the signal-to-noise ratio (SNR) of a tomogram is very low, making the structural recovery of macromolecular complexes extremely difficult. Another main image distortion is the missing-wedge effect. Since the cell sample can only be imaged at a limited tilt angle range (such as $\pm$ 60$^{\circ}$ ) to prevent excessive electron beam damage, the missing imaging wedge will cause structural distortions in the reconstructed tomogram. To reduce the influence of image distortions on the final recovered macromolecular complex structure, computational analysis methods for tomogram denoising, detection of subtomograms containing macromolecules, and subtomogram classification and segmentation are needed.

Another major problem is that a large number of cellular structures are still unknown to the scientific community \cite{han2009survey}. This will invalidate many reference-guided structural recoveries and pattern mining methods because if the unknown structure exists, prior knowledge such as resolved structures cannot be easily applied. Therefore, template-free methods designed specifically for cryo-ET analysis are needed. These include but not limited to discovering abundant and representative features automatically, and template-free averaging and classification of subtomograms.

Motivated by the need for computational methods designed for various cryo-ET data analysis tasks, we and collaborators \cite{aitom2019collatorators} aim to develop an open-source artificial intelligence platform, AITom, for performing large-scale cryo-ET data analysis pipelines. AITom is originated from the open-source software Tomominer \cite{frazier2017tomominer} and the open-source \textit{de novo} Multi-Pattern Pursuit (MPP) framework \cite{xu2019novo} developped at Prof. Frank Alber's lab \cite{alber2008lab}. AITom inputs a reconstructed 3D tomogram and computes a structural map with detected and recovered structures localized. We provide comprehensive tutorials to guide the users through each module. Providing the correct input format, each individual module can be executed separately. AITom also provides a hybrid of scripting and basic graphical user interface that enables users to perform interactive analysis on a remote server.

\begin{figure}
    \centering
    \includegraphics[width=0.94\textwidth]{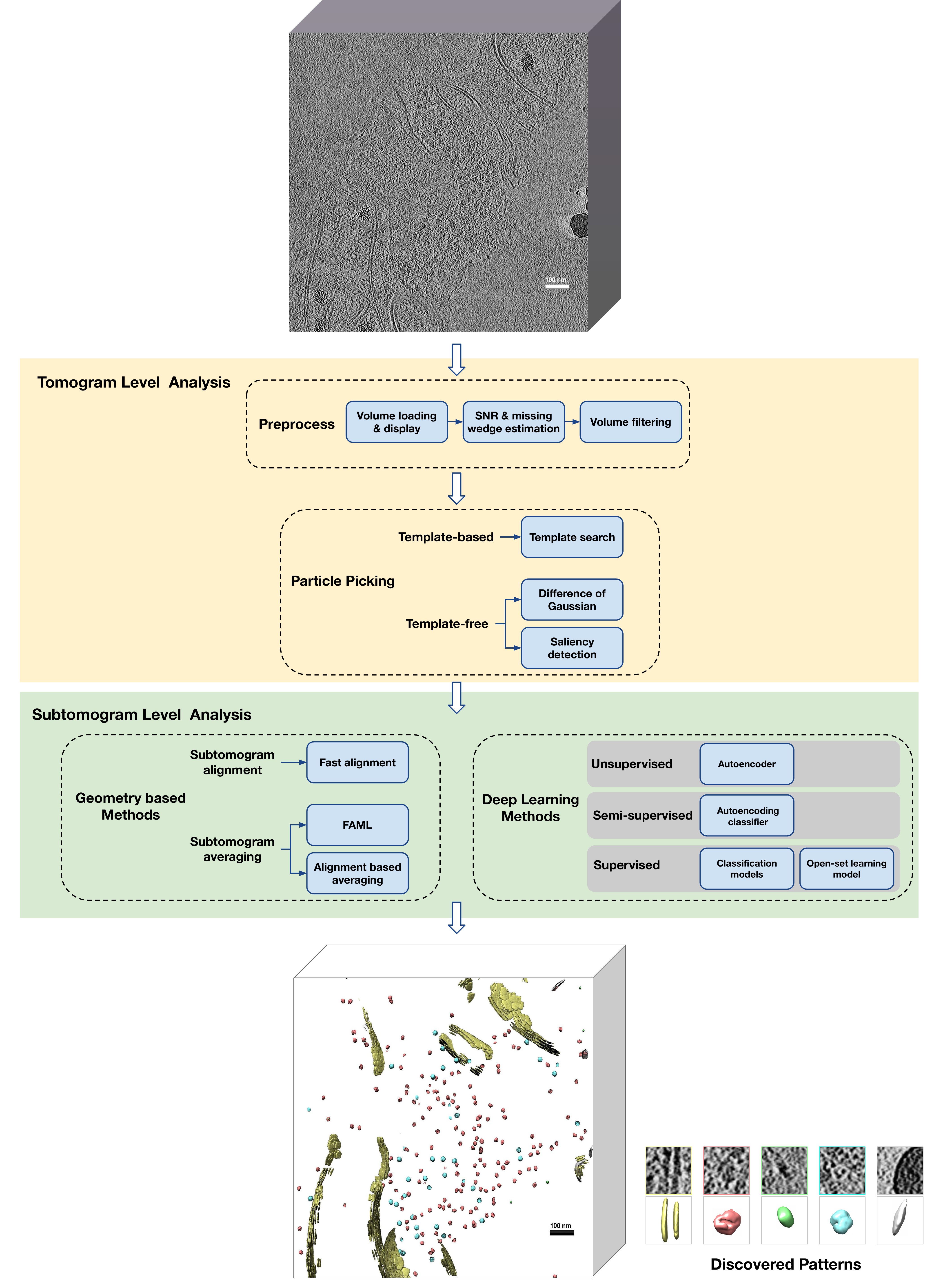}
    \caption{AITom workflow. AITom is demonstrated without using known structural templates on a tomogram of primary rat neuron culture \cite{guo2018situ} and recovered patterns (from left to right): mitochondrial membrane, ellipsoid of strong signals, borders of ice crystal, TRiC-like pattern, ribosome-like pattern, and their spatial distribution.}
    \label{fig:my_label}
\end{figure}

\section{Software implementation}
AITom implements a number of programs ranging from tomogram level preprocessing and particle picking to subtomogram level geometrical methods and deep learning methods (Figure \ref{fig:my_label}). Most of the programs are implemented in Python/C++. The deep learning methods are implemented using Keras backend by Tensorflow. Parallel computation using multiple CPU cores are provided for programs with high computation costs. 

\section{AITom analysis programs}

\subsection{Preprocess}
\subsubsection{Volume loading and displaying}
The input tomogram should be an MRC file. The visualization is enabled by using software IMOD \cite{kremer1996computer} and Python package matplotlib.

\subsubsection{SNR and missing wedge estimation}
The two major image distortion noise and missing wedge effect can be estimated to assist downstream analysis and parameter selection. AITom estimates the SNR (defined in \cite{frank2006three}) using the cross-correlation between pairs of aligned subtomograms containing the same structure. AITom estimates the missing wedge region using the cross-correlation between a missing wedge mask and the tomogram data projected to Fourier space.

\subsubsection{Volume filtering}
AITom implements two volume filtering methods, Gaussian filtering and anisotropic diffusion \cite{fernandez2003improved} for denoising the tomogram. The denoising preprocess step will enhance the structural signal for better downstream analysis.

\subsection{Particle picking}
Given a 3D tomogram, small subvolumes (subtomograms) need to be extracted to analyze the individual structures. This is known as particle picking. Manually picking subvolumes likely to contain macromolecular complexes requires long hours of labor by biologists. Furthermore, manual picking is biased by the prior knowledge of different biologists. Therefore, the automated algorithmic approach is preferred. Particle picking is divided into two general categories: (1) template-based particle picking matches a known structural template at a certain location to determine whether there is such a structure. Template-based particle picking is useful when certain structures need to be detected to analyze its spatial distribution; (2) template-free particle picking extracts subtomograms from potential structural regions without relying on known structure templates. Template-free particle picking is useful when systematic detection of both known and unknown structures is needed.

\subsubsection{Template matching}
Template matching is a popular template-based particle picking method by calculating the cross-correlation between a structural template and a subvolume at a certain location \cite{bohm2000toward}. Since a particle can be of random orientation at a random location inside the tomogram, template matching first generates possible orientations of the structural template by rotating it with uniformly sampled degrees. Then the cross-correlation between a copy of the structural template and all tomogram locations is computed by convolution. The value of the cross-correlation map between a structural template and a tomogram is determined by the highest cross-correlation value among all copies of the structural template. A manually selected cross-correlation threshold is needed to determine whether a structure exists at a certain location. The template matching implemented in AITom  performs convolution by the fast Fourier transform method \cite{nussbaumer2012fast}.

\subsubsection{Difference of Gaussian}
\label{sec:dog}
Difference of Gaussians (DoG), a fast approximation of Laplacian  of Gaussian (LoG), is widely used for particle picking in single-particle cryo-EM analysis. We extended DoG to 3D cryo-ET analysis \cite{pei2016simulating}. DoG detects peaks in the subtraction of two Gaussian filtered image with different standard deviations as potential particles. Local density peaks are filtered based on a threshold to exclude results likely to be from noise and based on the pair-wise distance to exclude multiple hits to one particle. 3D DoG is implemented in AITom with parameters such as smoothing factors and the minimum distance between detected locations to tune. Tomograms with detected particle locations annotated can be visualized in AITom using software IMOD \cite{kremer1996computer} to qualitatively assess the particle picking performance to assist parameter tuning. 

\subsubsection{Saliency detection}
Saliency detection is a cutting-edge machine learning approach for inferring the likelihood of an image subregion stands out relative to its background \cite{zhu2014saliency, qin2015saliency}. We extended saliency detection to a cryo-ET template-free particle picking framework in \cite{zhou2018feature}, which is based on matrix decomposition of features extracted from super-voxels. The feature matrix is decomposed by the Robust-PCA method \cite{candes2011robust} to a low-rank matrix representing background information and a saliency matrix representing structural information. This saliency detection framework is implemented in AITom as a particle picking method as well as a tomogram denoising method.

\subsection{Geometrical methods}
Template-free cryo-ET analysis relies heavily on geometrical methods. AITom implements geometrical methods for two important tasks for structural classification and recovery: (1) Subtomogram alignment which aligns a structural template and a subtomogram or a pair of subtomograms; (2) Subtomogram averaging which estimates the underlying structure from multiple copies of noisy subtomograms.

\subsubsection{Fast alignment}
Subtmogram alignment estimates the 3D rigid body geometric correspondence between a structural template and a subtomogram or a pair of subtomograms. AITom implements three fast subtomogram alignment methods \cite{xu2012high, xu2013high, lu2019fine}. Compared with exhaustive search based methods, fast subtomogram alignment methods apply heuristics to perform this computationally intensive task efficiently. Providing the known missing wedge mask, the missing wedge effect is taken into account by using the constrained cross-correlation \cite{forster2008classification}. 
\subsubsection{Alignment based averaging}
To eliminate potential bias, subtomogram averaging is often done without any external structural templates. Subtomogram alignment can be extended to alignment based subtomogram averaging \cite{briggs2013structural}, by aligning a set of subtomograms containing homogeneous structures to their simple average and re-averaged for the next iteration. The resolution of the subtomogram average will gradually improve during the iterative process until convergence. When the set of subtomograms contains heterogeneous structures, subtomogram averaging and classification divides the set into clusters of homogeneous structures and computes their averages. To achieve the simultaneous clustering and averaging, AITom implements a framework consisting of dimension reduction, clustering, and optimal cluster cut-off selection to integrate the clustering process into the iterative process \cite{xu2012high}.      

\subsubsection{Fast alignment maximum likelihood averaging}
Fast Alignment Maximum Likelihood (FAML)  \cite{zhao2018integration} integrates a fast alignment method and the maximum likelihood based averaging approach for fast and robust subtomogram averaging. The fast alignment method \cite{xu2012high}, computes a set of suboptimal rigid transformations under a translation-invariant upper-bound. The maximum-likelihood based approach \cite{scheres2009averaging}, defines a data model and derives an EM algorithm for updating parameters. The FAML algorithm significantly improved the recovered structure resolution as compared with the fast alignment based approach, and achieved 2 to 5 times speedup compared to the maximum-likelihood based approach without loss in resolution.

\subsection{Deep learning methods}
Since 2017, deep learning methods have been proposed for cryo-ET data analysis and attracted a lot of attention \cite{gubins2019classification}. Compared with traditional methods, deep learning methods have several advantages including significantly faster prediction and progressively improved model performance with big data. Deep learning methods have been applied to subtomogram semantic segmentation, classification, and clustering. AITom implements several deep learning methods ranging from supervised, semi-supervised, to unsupervised training modes. All the model architectures and hyperparameters can be easily modified and tuned.  

\subsubsection{Classification models}
AITom implements three deep learning classification models for cryo-ET subtomogram data \cite{xu2017deep, che2018improved}. The three models are: 1) DSRF3D-v2 (Deep Small Receptive Field version 2:  a 3D variant VGGNet \cite{simonyan2014very} based Convolutional Neural Network (CNN) architecture featured with sequentially deep stacked layers and small 3D convolution filters with size of 3x3x3. 70 \% dropout is used for generalization purposes. 2) RB3D (Residual Block 3D): a 3D variant residual block \cite{he2016deep} based CNN model with four bottleneck residual blocks are connected sequentially. For each block, two paths merge together at the end of the block.  One path contains one 1x1 layer to reduce dimension, a 3x3 layer and a 1x1 layer for restoring dimension.  The other path, considered as a``shortcut”, only contains a 3x3 convolutional layer. 50 \% dropout is used for generalization purposes. 3) CB3D: a model based on C3D \cite{tran2014c3d},  originally proposed  to  be  trained  on  large  scale  supervised  video  datasets. This model treats a 3D particle as a continuously changing 2D object.  Five max pooling layers are mixed among the convolutional layers and 50 \% dropout is applied at the end for generalization purpose.

In addition, for supervised CNN classification models, it is important to obtain training data with labels to train the model. However, for cryo-ET subtomograms, labeled data is not always available because it requires extensive manual labeling. It is ideal to use simulated datasets to train the model and predict on experimental datasets. However, due to the domain shift (simulated data and experimental data has different noise patterns), the prediction accuracy is limited. AITom implements an adversarial domain adaptation technique to solve this issue \cite{lin2019adversarial}. The domain adaptation consists of three stages: 1) Train the classification network using simulated data, 2) Confuse the Domain Discriminator between features from the simulated training data and features from the testing data, 3) Predict on the testing data using the indiscriminable features from the extractor.

\subsubsection{Autoencoder}
For unsupervised learning from cryo-ET data, AITom implements a convolutional autoencoder that coarsely groups and filters raw subtomograms  \cite{zeng2018convolutional}. Since a large portion of raw subtomograms contains uninteresting structures or pure noise, biologists usually need to manually select interesting subtomograms from tens of thousands of subtomograms. The autoencoder reduces this extremely laborious task to the selection among less than 100 clusters with relatively homogeneous structures. To facilitate grouping together the subtomograms containing the same structure but in different orientations, this method contains a pose normalization preprocessing step to normalize the orientation and displacement of structures inside a subtomogram. K-means clustering algorithm is applied on the latent-space encodings of subtomograms for clustering. 

\subsubsection{Autoencoding classifier}
When both labeled and unlabeled subtomogram data is available, semi-supervised learning can improve both the supervised classification performance and the unsupervised encoding learning performance. AITom implements a semi-supervised autoencoding classifier integrating the above classification model and the autoencoder to simultaneously learn both of them. The feature extraction layers are shared to mutually reinforce the two tasks \cite{liusemi}.

\subsubsection{Open-set learning model}
In an open-set recognition scenario, the testing data contains novel structural classes unseen in the training data. AITom implements an open-set learning model with specifically designed losses to enforce extracted features with a large inter-class variance and a low intra-class variance \cite{duopen}. Then the testing data can be clustered using the extracted features from the open-set learning model.

\subsection{Remote analysis with scripting and graphical hybrid}

The fast accumulation of cryo-ET data makes it difficult for a research lab to store and process the data using only a single workstation. Therefore, storing and processing cryo-ET on servers or computer clusters shared by multiple users or even multiple labs is preferred. Nevertheless, in practice cryo-ET analysis often involves graphical user interface (GUI) for interactive analysis. On the other hand, due to limited data transfer rate, a fully featured and smooth GUI between a client computer and a remote server is often difficult to implement. To solve this problem, we have Jupyter Notebook based approach that allows the users to perform cryo-ET data processing and analysis using both python scripts and basic graphical interaction. 
After connecting to the server remotely, users can easily run python scripts and view the results in the local browser.  

\subsubsection{Example use cases}
\begin{figure}
  \centering
  \includegraphics[width=0.6\textwidth]{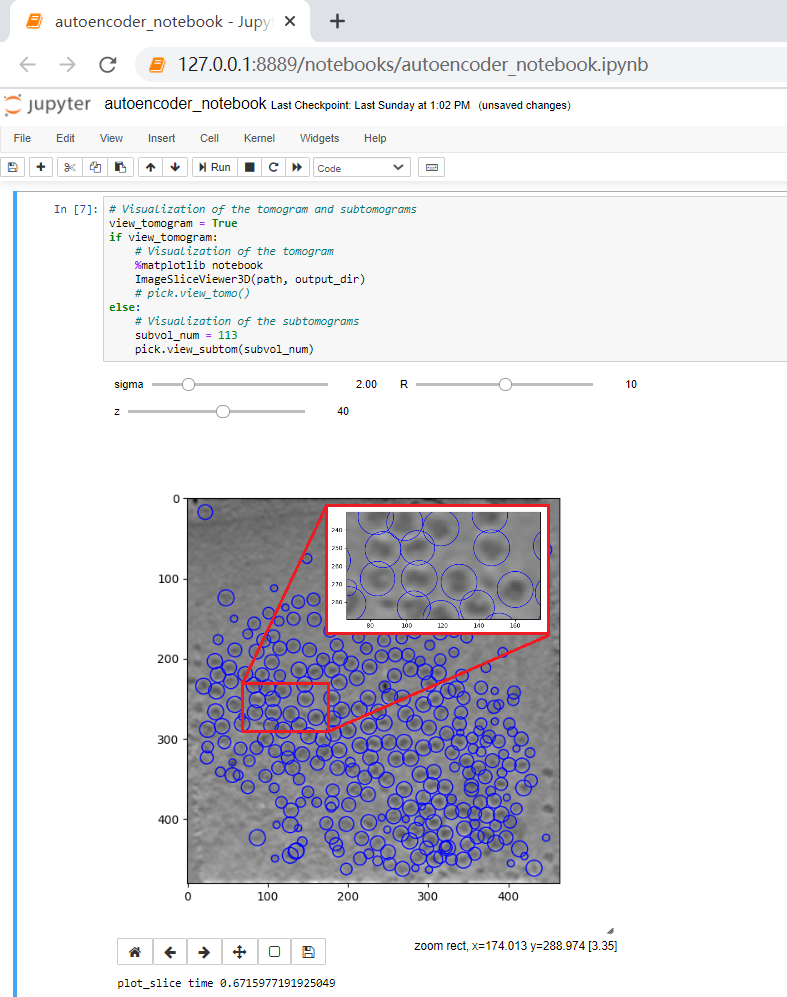}
  \caption{Overview of particle picking results. Particles are circled by blue circels. 'sigma' controls gaussian kernel size for denoising, 'R' controls the radius of blue circles and 'z' controls which slice to display. Part of the slice can be zoomed in for better visualization.}
  \label{fig:gui_1}
\end{figure}

In this part, we explain how to conveniently use AITom to do particle picking, classification and reconstruction. AITom provides interfaces for some modules such as particle picking and autoencoder. Users can pass in parameters in the terminal or Jupyter Notebook to quickly obtain results. As described in Section \ref{sec:dog}, DoG \cite{pei2016simulating} is widely used for particle picking in single-particle tomogram, but a few parameters need to tune. To offer fast visualization, AITom provides a 2D display window with some buttons and scroll bars. In practice, sigma of Gaussian filters is the most important parameter to tune, which should be approximately equal to the size of the particles to be picked. After particle picking, picked particles of each slice are circled by blue circles (Figure \ref{fig:gui_1}). Users can first try to use the automatically selected parameters and decide whether to tune the parameters or go to the next step after viewing the overview results. Next, manual selection is an option. Users can manually determine which particles to keep for further research. At this point, particle picking is complete. The results will be saved in a pickle file, and each particle center corresponds to a UUID. Autoencoder is then used for classification and reconstruction. By default, 32x32x32 subtomograms are taken from the tomogram as input. Since \cite{liusemi} is an unsupervised learning method and k-means is used for classification, the number of cluster centers needs to be set as an parameter equal to the number of particle types. For the reconstruction task only, simply set the number of cluster centers to 1. After the training process is completed, the reconstruction particles will be displayed in slices.

\section{Discussion}
AITom performs large scale cryo-ET data analysis. Besides traditional image processing methods, AITom implements many state-of-the-art deep learning models specially designed for cryo-ET data. AITom could be used to complement many other cryo-ET analysis software [e.g. \citealp{han2017autom, kremer1996computer, scheres2012relion, himes2018emclarity, tang2007eman2, hrabe2012pytom, castano2012dynamo, de2016scipion}]. 
Open-sourcing our initial code for AITom is only the very first step. We plan to develop AITom as a collaborative project to continue to disseminate new cryo-ET analysis algorithms and deep learning models to directly benefit the cryo-ET community. We also intend to interface AITom with other software. We welcome researchers to apply AITom and provide feedback to us. We welcome collaborators to join the development of AITom.

\section{Acknowledgement}

This work was supported in part by U.S. National Institutes of Health (NIH) grant P41GM103712, and R01GM134020, U.S. National Science Foundation (NSF) grants DBI-1949629 and IIS-2007595. XZ was supported by a fellowship from Carnegie Mellon University's Center for Machine Learning and Health.


\newpage
\bibliographystyle{unsrt}
\bibliography{references}

\end{document}